\documentclass[prb,aps,twocolumn,showpacs,home,floats]{revtex4}

\usepackage{graphics}
\begin{document}

\title{ {\rm\small\hfill (submitted to Phys. Rev. B})\\
Insights into the function of silver as an oxidation catalyst
by \\
ab initio, atomistic thermodynamics}

\author{Wei-Xue Li,$^{1}$ Catherine Stampfl,$^{1,2}$ and 
Matthias Scheffler$^{1}$}
\affiliation{$^{1}$Fritz-Haber-Institut der Max-Planck-Gesellschaft,
Faradayweg 4-6, D-14195 Berlin-Dahlem, Germany\\
$^{2}$School of Physics, The University of Sydney, Sydney 2006 Australia}

\date{Received 24 April 2003} 

\begin{abstract}
To help understand the high activity of silver as an oxidation
catalyst, e.g., for the 
oxidation of ethylene to epoxide and the dehydrogenation of methanol to
formaldehyde, 
the interaction and stability of many different oxygen species at the Ag(111) 
surface has been studied for a wide range of coverages. Through calculation
of the free energy, as obtained from density-functional theory and
taking into account
the temperature and pressure via the oxygen chemical 
potential, we obtain the 
phase diagram of O/Ag(111).
Our results reveal that a {\em thin}  surface-oxide
structure is most stable for the temperature 
and pressure range of ethylene epoxidation 
and we propose it (and possibly other similar structures) contains the
species actuating  the catalysis. 
For higher temperatures, low coverages of chemisorbed
oxygen are most stable, which could also play a role in oxidation reactions.
For temperatures greater than about 775~K
there are {\em no} stable oxygen species, except
for the possibility of
O atoms adsorbed at under-coordinated surface sites (i.e.,
imperfections, defects).
At low temperatures ($\lesssim$ 400~K at atmospheric pressure),
provided kinetic limitations can be overcome,
thicker oxide-like structures are predicted. 
Due to their low thermal stability, however,
they can be ruled out as playing an 
important role in the heterogeneous reactions under technical conditions. 
Bulk dissolved oxygen and a molecular ozone-like species adsorbed at a surface
vacancy, as have been proposed in the literature, are 
found to be energetically unfavorable.
The employed theoretical approach
for calculating {\em free energies} and predicting
the lowest energy structures in contact with species in the
environment (``{\em ab initio}, atomistic thermodynamics''), 
affords investigation of a system that seemlessly connects
standard ($T=0$~K) DFT results, characteristic of ``typical'' theoretical
surface science studies, through to those valid for the conditions
of catalysis.
Though the error bar of the noted theoretical temperatures is noticeable
($\pm \approx$55~K), the identified trends and physical descriptions
are useful.
\end{abstract}


\maketitle

\section{Introduction}  
The identification of
active species for desired chemical reactions is a main
quest in the field of heterogeneous catalysis.~\cite{ertl} Whether a species
is active or not depends on many factors, including the
nature of its interaction with
the catalyst material, as well as the temperature and pressure at which
the catalytic reactions
take place. The main approach for
gaining quantitative, microscopic understanding of
various properties and processes, has been to perform experiments under
standard surface science conditions, i.e.,  under ultra high vacuum
(UHV) and relatively low (room or lower) temperatures,
with the hope that what is learnt may be relevant and/or
extrapolated to the system under
the ``real'' catalytic conditions of high temperatures and pressures.
Theoretical studies of adsorption and reactions at surfaces
have, at best, 
been carried out by first-principles approaches,
which, while highly accurate, typically do not take into account the temperature
or
the gas-phase environment with which the surface is in contact.
A recent focus, both experimentally and theoretically, is to
attempt to study systems which more closely resemble those
of true catalytic ones.~\cite{surfsci-500}  Experimentally, this is done
by performing 
{\em in situ} real-time measurements carried out under
high temperature and pressure conditions, and by the systematic study of,
e.g., well-characterized metal clusters deposited
on support materials and chemical reactions thereon (see
e.g., Refs.~\onlinecite{peden,somorjai,somor2,freund,ferrer,frenken}).
Such studies are important in particular for understanding
those systems which exhibit a 
``pressure- and/or materials-gap'', i.e., where different behavior
is exhibited under low and high pressures,
or when results for model single-metal catalysts do not represent
those of the ``real'' catalyst, typically consisting of a
dispersed metal on a support material.
Often the pressure- and materials-gaps are related and occur 
simultaneously.

The oxygen--silver system is an example which exhibits a pronounced
pressure-dependence:
Despite its unique activity as {\em the} catalyst for ethylene 
epoxidation 
(C$_{2}$H$_{4} + \frac{1}{2}$O$_{2}\rightleftharpoons$ C$_{2}$H$_{4}$O,
performed at 500-600~K and atmospheric pressure) and partial 
oxidation of methanol to formaldehyde 
(CH$_{3}$OH+$\frac{1}{2}$O$_{2} \rightleftharpoons$ CH$_{2}$O+
H$_{2}$O - ``oxydehydrogenation'', or by direct dehydrogenation,
CH$_{3}$OH $\rightleftharpoons$ CH$_{2}$O+H$_{2}$, performed
at 700-900~K and atmospheric pressure), 
these reactions do not occur with any significant probability
under UHV conditions.~\cite{sant87,rocc01,savi02} 
This behavior of the oxygen--silver system might
be related to a materials-gap, because
for studies of the partial oxidation of methanol,
drastic differences are observed in the performance for fresh
silver catalysts and those which have been exposed to high temperature
($T>873$~K) and pressure (atmospheric) conditions. The temperature dependence of
the conversion of the clean Ag material to the active
catalyst exhibits a pronounced hysteresis,
i.e., the conversion is different when raising the temperature from
low to high, compared to from high to low.
This was the case 
for high temperatures up to 873~K, as well as for lower temperatures
up to 560~K and 750~K. 
These hysteresis effects were 
shown from scanning electron microscopy (SEM)
studies to be related to 
significant morphological changes which occur in the catalyst 
depending on the temperature, and that do not revert to the original
structure
when lowering the temperature, i.e., the changes
are irreversible.~\cite{nagy}
Structural changes in silver also occur when 
in contact with a {\em pure} O$_{2}$ environment
at high temperatures ($\sim$900~K) and atmospheric pressure, namely,
facets with the Ag(111) orientation result.~\cite{nagy,bao} 
This is presumably because under these conditions 
it is unlikely that there will be any oxygen on the surface and
the (111) surface of Ag has the lowest surface energy.

It has been suggested that  
the elusive sub-surface or ``bulk-dissolved'' oxygen species
play an essential role in the  above-mentioned catalytic
reactions, which so far have escaped a precise 
characterization.~\cite{nagy,bao,gran85,van86,bukh94,herein-96}
In addition to this aspect,
there are other controversial debates in the literature
concerning the nature of the
active species, e.g.,  atomic versus molecular (ozone-like)
oxygen.~\cite{bukh011,avde01}  
Another puzzling issue is how silver, as a noble metal, can function
as a good catalyst when  it only binds adparticles relatively
weakly on the surface compared to the transition-metal catalysts to
the left of it in the periodic table.
Catalysis clearly involves a number of complex processes, namely the
dissociation of molecules and the creation of chemically active
species, and subsequent interaction
and reaction between the particles to form the product, which 
desorbs from the surface. 
Furthermore, it is possible (maybe likely) that some fragments
of the reactant molecules (e.g., H or C) modify the catalyst material
and thereby participate in the creation of the active species.
In the present paper we tackle only part of the problem,
namely, the study of the interaction of silver with oxygen
in order to identify and exclude  possible active O species for the
above-mentioned heterogeneous catalytic reactions.
This is an essential prerequisite for                           
modeling the full catalytic process. 

Although we study explicitly the (111) surface of silver,
our general findings 
are expected to be relevant for silver {\em per se}, and possibly
to have implications for gold,
also a noble metal oxidation catalyst~\cite{gold}
which, for the (111) surface, exhibits a restructuring
and chemisorption of oxygen atoms at elevated temperatures
(500--800~K) and
atmospheric pressure,~\cite{gold1,gold2}
similar to Ag(111).
They may also relate to copper,
which catalyses the oxidation of methanol to formaldehyde.~\cite{copper}

In our previous publications,~\cite{wxli01,wxli02} 
the interaction between oxygen
and Ag(111) was thoroughly studied by density-functional
theory (DFT). In these works we
considered almost all possibly relevant oxygen species,
namely, pure on-surface
chemisorbed oxygen, pure sub-surface and bulk-dissolved
oxygen, surface- and bulk-substitutional adsorption,
structures involving both oxygen on and under the top Ag layer, 
as well as surface-oxides
and a molecular (ozone-like) species adsorbed at a surface vacancy. 
We studied these oxygen species for a wide range of coverages where
the following results were obtained: 
At low coverages oxygen atoms adsorb on the surface. With increasing
coverages, however, strong repulsive interactions develop between
the partially negatively charged O atoms and 
the adatoms are driven to penetrate into the sub-surface region, and/or
to give rise to a reconstructed,
thin surface-oxide layer, when the coverage 
is larger than $\sim$0.25~ML. 
For higher oxygen contents, thicker oxide-like films
are predicted to form.

We also found that the presence of sub-surface oxygen when
bonded to the same surface Ag atoms as an oxygen atom 
on the surface,
dramatically 
modifies the electronic properties of the on-surface oxygen 
and vice-versa. 
Depending on the particular geometry, it can 
give rise to a notable stabilization or destabilization, but the
energetically 
favorable structures induce a {\em stabilization} and a lowering of the
density of states at the Fermi level.
In contrast to the conclusions concerning the crucial role
of bulk-dissolved oxygen based
on interpretations of experimental results,
we found that this species
is energetically {\em unfavorable}, as  is
bulk- and surface-substitutional oxygen,
as well as the molecularly adsorbed ozone-like species at a vacancy,
when the system is in thermal equilibrium.~\cite{wxli01,wxli02}
On the other hand, the binding energy of atomic oxygen adsorbed
at under-coordinated silver  atoms
is considerably stronger than on the terraces, but the concentration
of such species obviously depends on the number of defects and 
the morphology of the substrate.

Nevertheless, temperature and pressure can affect the stability
of structures and this has not been
investigated so far. Clearly this is important for gaining
an understanding of the system under catalytic conditions and
this is the main topic of the present paper. 
Some of these results were reported briefly in Ref.~\onlinecite{wxli03}.
Our study uses the approach of  ``{\em ab initio}, atomistic
thermodynamics", that
is, performing systematic DFT calculations
for all O species
that could conceivably be relevant, and take into account the effect of
the environment, which we do via the pressure and temperature
dependence of the oxygen chemical potential
(see, e.g., Refs.~\onlinecite{Weinert86,Scheffler87,Kaxiras87,Qian88,wang98,karsten01}).
From the resulting free energies (obtained using our
previously obtained energetics~\cite{wxli01,wxli02}),
we derive the phase diagram of oxygen at the
silver (111) surface.
Since our earlier 
studies~\cite{wxli01,wxli02} predict that oxide-like structures
form for high O concentrations,
we also investigate the surface and bulk properties of 
the most stable bulk oxide, namely, di-silver oxide, Ag$_{2}$O. 

The paper is organized as follows: The calculation procedure, as well as the
thermodynamical method are explained in Sec.~II, 
and in Sec.~III results are presented for the bulk properties of Ag$_2$O.
In Sec.~IV, 
we investigate various terminations of the Ag$_2$O(111) surface 
as a function of the oxygen chemical
potential. 
Section~V describes investigations into the transition from oxide-like
structures (where the positions of the Ag atoms
are commensurate with those of the Ag(111) substrate)
to the true oxide film which is laterally expanded in comparison.
In Sec.~VI, the pressure--temperature phase 
diagram of the oxygen--silver system is presented 
and compared to experimental results. 
Finally, the conclusions are given in Sec.~VII. 

\section{Calculation Method \label{sec:cal}}

The DFT total-energy calculations are
performed using the pseudopotential plane wave method~\cite{fhi98} 
within the generalized gradient approximation (GGA).~\cite{pbe96,white94} 
The pseudopotentials are generated by the scheme of Troullier and Martins
with the same functional.~\cite{fuch99,trou91} The wave functions are expanded
in plane waves with an energy cutoff of 50~Ry. In the $(1\times1)$ 
surface unit cell, 
corresponding to the periodicity of clean Ag(111),
21 special {\bf k}-points are used in the surface irreducible
Brillouin zone (IBZ) for the Brillouin-zone integration.~\cite{cunningham} 
Equivalent {\bf k}-points to these are used for all of the surface structures
studied for consistency, i.e., to maximize the accuracy when comparing
the energetics of different coverages as calculated in different supercells.
We employ
a Fermi function with a temperature broadening parameter of
$k_{\rm B} T^{\rm el}=0.1$~eV to improve the convergence, and the total energy is
extrapolated to zero temperature. 
($k_{\rm B}$ is the Boltzmann constant.)
The structures 
are created on one side of the slab and 
the resulting induced dipole moment is taken into account by
applying a dipole correction.~\cite{jorg92}
For the adsorption structures on Ag(111), we use five layers of Ag to
model the surface and
typically keep the bottom two or three layers fixed at the bulk-like positions
and fully relax  all the other atoms
until the forces on the atoms are less than 0.015~eV/\AA\,. 
The oxide surfaces of
Ag$_2$O(111) are simulated by symmetrical 
slabs (i.e., with inversion symmetry). 
We tested thicknesses corresponding to
five, seven, and nine metal layers, separated by 15~\AA\, of vacuum. 
Our results show that 
the systems can be safely described using five metal layers with inversion
symmetry.

In {\em ab initio} theory, the consideration of high temperature and
high pressure can be achieved by explicitly taking into account the 
surrounding gas phase in terms of ``{\em ab initio}, atomistic
thermodynamics'' (cf.
e.g. Refs.~\onlinecite{Weinert86,Scheffler87,Kaxiras87,Qian88,wang98,karsten01}).
This is also an appropriate (first) approach to steady-state catalysis,
which is often run close to thermodynamic equilibrium (or a constrained 
equilibrium) to prevent catalyst degradation. 
In the following we outline how
the combination of thermodynamics and DFT
can be applied to obtain the lowest-energy surface structures 
with a surrounding gas phase,
thus enabling us to construct a $(T,p)$-diagram of the stability
(or metastability) regions of different surface phases.

We consider a
surface in contact with an oxygen atmosphere which is described
by an oxygen pressure $p$ and a temperature $T$.
The environment then acts as a reservoir, as it can give (or take)
oxygen to (or from) the substrate without changing the 
temperature or pressure.
We calculate the surface free energy,
\begin{eqnarray}
\gamma(T,p)=(G - N_{\rm Ag} \mu_{\rm Ag} 
- N_{\rm O}  \mu_{\rm O}) / A \quad ,
\label{eq-gibbs}
\end{eqnarray}
\noindent
where $N_{\rm O}$ and $N_{\rm Ag}$ are the number of oxygen and
silver atoms, and $A$ is surface area.
The $T$ and $p$ dependence is mainly given by $\mu_{\rm O}$, the
oxygen chemical potential, i.e., by the O$_{2}$ gas phase atmosphere:
\begin{eqnarray}
\mu_{\rm O}(T,p) = 1/2[E_{\rm O_{2}}^{\rm total} + 
\tilde{\mu}_{\rm O_2}(T,p^{0})
 \;+\;  \;k_{B} T
 \;{\rm ln} \left( \frac{p_{\rm O_2}}{p^{0}} \right)],
\label{eq:ochem}
\end{eqnarray}
where $p^{0}$ corresponds to atmospheric pressure and
$\tilde{\mu}_{\rm O_2}(T,p^{0})$ includes the contribution from
rotations and vibrations of the molecule, as well as the
ideal-gas entropy at 1 atmosphere. It can be calculated or taken
from experimental values from thermodynamic tables -- the
difference between theory and experimental values is marginal in the 
temperature range of interest.
The silver chemical potential is
\begin{eqnarray}
\mu_{\rm Ag}= \left\{ 
  \begin{array}{l}
      g_{\rm Ag-bulk},   \mbox{for adsorption on a silver substrate}\\ 
      \frac{1}{2}(g_{\rm Ag_{2}O-bulk} - \mu_{\rm O}),  \mbox{when
      bulk Ag$_{2}$O is present}
   \end{array}
   \right.
\end{eqnarray}
We use small symbols ($g_{\rm Ag-bulk}$, $g_{\rm Ag_{2}O-bulk}$) in
Eq.~3 to 
indicate that these are free energies per Ag atom (top) and per Ag$_{2}$O
unit (bottom). These two quantities are calculated by DFT-GGA.
In Eq.~1, we have furthermore,
\begin{eqnarray}
G = \left\{
  \begin{array}{l}
   G^{\rm slab}, \mbox{to obtain surface energies}\\
   G_{\rm O/Ag(111)}^{\rm slab} - G_{\rm Ag(111)}^{\rm slab},
    \mbox{to obtain adsorption} \\
   \mbox{\hspace{3.5cm} energies on Ag(111)}
   \end{array}
   \right.
\end{eqnarray}
These quantities are also calculated by DFT-GGA. For the temperature 
dependence of $\tilde{\mu}_{\rm O_{2}}(T,p^{0})$, we use thermodynamic tables
[Ref.~\onlinecite{janaf}] (i.e., experimental data), while 
the $T=0$~K value of Eq.~\ref{eq:ochem} is 
\begin{displaymath}
\mu_{\rm O}(T=0{\rm K},p) = \frac{1}{2}E_{\rm O_{2}}^{\rm total} 
\end{displaymath}
\begin{equation}
\simeq  E_{\rm Ag_{2}O-bulk}^{\rm total} \; - \;2 
E_{\rm Ag-bulk}^{\rm total}
\; + \; H^{f}_{\rm Ag_{2}O-bulk} 
\label{eq5}
\end{equation}
In the above equation we use the experimental value of the heat of formation,
$H^{f}_{\rm Ag_{2}O-bulk}$, because of its
very small value (0.323~eV at the standard state (room
temperature and atmospheric pressure),~\cite{crc}
or 0.325~eV extrapolated back to $T\rightarrow 0$~K), which implies
that the error introduced in the later discussion by using
the theoretical value (0.092~eV) would be significant.
We note that the error in $H^{f}_{\rm Ag_{2}O-bulk}$
is due to the significant overbinding of O$_{2}$, and
the underbinding obtained for bulk Ag and bulk Ag$_{2}$O. 

We choose the zero reference state of $\mu_{\rm O}(T,p)$ 
to be the total energy of oxygen in an isolated molecule, i.e., 
$\mu_{\rm O}(0{\rm K},p)=\frac{1}{2}E^{\rm total}_{\rm O_{2}} \equiv 0$.
This will be called the ``oxygen-rich'' condition.
The so-called ``oxygen-poor'' value of $\mu_{\rm O}$ is taken as, 
\begin{equation}
\mu_{\rm O}^{\rm poor}\; = \; E^{\rm total}_{\rm Ag_{2}O-bulk}
\;-\; 2 \; E^{\rm total}_{\rm Ag-bulk} \; .
\end{equation}
\noindent
If the oxygen chemical potential becomes lower 
(more negative) than $\mu_{\rm O}^{\rm poor}$,
di-silver oxide starts to decompose. For oxides
of different transition metals, the
``oxygen poor'' limit can vary significantly depending on the heat of formation. 
For example, 
what may be rather ``oxygen rich'' for RuO$_{2}$ (with
$H_{f}=1.60$~eV) is
already too ``oxygen poor'' for Ag$_{2}$O to be stable. 
The typical range of the allowed oxygen chemical potential is 
determined and limited 
by the pressures and temperatures used in industry and
laboratories, namely, from several 
to thousand Kelvin, and from ultra high vacuum ($<10^{-12}$ atm.)
to several hundreds of 
atmospheres. 

For 
our investigations into various terminations of Ag$_{2}$O(111) surfaces,
we will also consider the energy required
to remove an oxygen atom from the surface and bring
it to the vacuum. We call this the ``removal energy'', and it is
defined as,
\begin{equation}
  E^{\rm removal}
  =-E_{\rm O/Ag_{2}O(111)}
  +(E_{\rm Ag_{2}O(111)}^{\rm ref}+E_{\rm O })
  \quad ,
  \label{eq:ebind}
\end{equation}
where $E_{\rm O/Ag_{2}O(111)}$ is the total energy
of the surface termination of interest, containing
the O atom to be removed, $E_{\rm Ag_{2}O(111)}^{\rm ref}$ is
that of
the reference system, i.e.,
the surface structure remaining after the O atom has been removed. 
$E_{\rm O}$ is the total energy of a free O atom where the
spin-polarization energy ($-$1.60~eV) is taken into account.
The removal energy reflects the bondstrength of the removed
O atom to the surface.

To help analyze the electronic structure,
we also calculate the density of states, $N(\epsilon)$, as well as 
the state-resolved DOS, or projected DOS, $N_{\alpha}(\epsilon)$.
The definitions are given in Ref.~\onlinecite{matthias}.

\section{Bulk A\lowercase{g}${\rm _2}$O \label{sec:ag2o}}

\begin{table}[t!]
\caption{Calculated DFT-GGA 
and experimental bulk properties of Ag$_{\rm 2}$O. The
  lattice constant, $a_{\rm 0}$, the distance between Ag atoms, 
  $d_{\rm Ag-Ag}$ and oxygen atoms, $d_{\rm O-O}$, and the bondlength
  between nearest neighbor oxygen and silver atoms, 
  $b_{\rm O-Ag}$, are given in \AA\,. The bulk modulus,  $B$, (in Mbar),
  and the heat of formation, $H^{f}_{\rm Ag_{2}O-bulk}$, 
  (in eV per oxygen atom) are
  listed as well.  \label{tab:ag2o}  }
\begin{tabular}{lllllll}
  & $a_{\rm 0}$    & $d_{\rm Ag-Ag}$ & $d_{\rm O-O}$ & $b_{\rm O-Ag}$ 
  & $B$     & $H^{f}_{\rm Ag_{2}O-bulk}$ \\ \hline
  Exp.   
  & 4.72$^{\rm a}$ & 3.34$^{\rm a}$ & 4.09$^{\rm a}$ 
  & 2.04$^{\rm a}$ & -           & 0.323$^{\rm c}$  \\
  & 4.74$^{\rm b}$ & 3.35$^{\rm b}$ & 4.10$^{\rm b}$ 
  & 2.05$^{\rm b}$ & -            & -  \\
  \hline
  Theory & 4.91         & 3.47           & 4.25          & 2.13
  & 0.647     & 0.092   
\end{tabular}
$^{\rm a}$ Reference \onlinecite{wyck64};~\ $^{\rm b}$ Reference
\onlinecite{well84};~\ $^{\rm c}$ Reference \onlinecite{crc}.
\label{tab1}
\end{table}

\begin{figure}
\scalebox{0.55}{\includegraphics{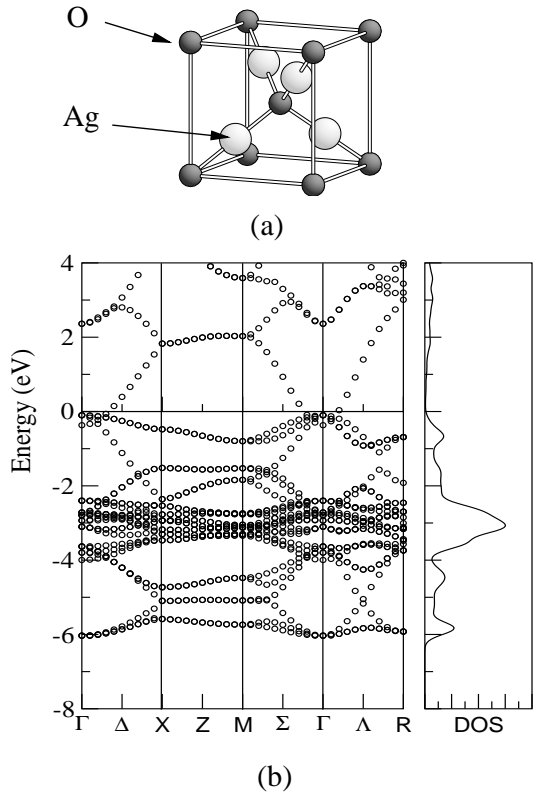}}
\caption{
(a) The cuprite structure of bulk Ag$_{\rm 2}$O. The oxygen and silver
  atoms are indicated by the dark and pale grey circles, respectively. 
(b) The band structure of Ag$_{\rm 2}$O calculated at the theoretical
  lattice constant (left), and the corresponding density of 
  states (DOS) is shown at the right.
  The energy zero corresponds to the Fermi level.}
\label{fig:band}
\end{figure}

The most stable silver oxide is Ag${\rm _2}$O, which decomposes at 
460~K under atmospheric pressure.~\cite{crc} The crystallographic structure 
of di-silver oxide, Ag$_{\rm 2}$O, is shown in Fig.~\ref{fig:band}a. It has 
the cuprite structure with six atoms per cubic unit cell, (two oxygen atoms and 
four silver atoms) where the oxygen atoms form a body centered cubic
lattice, while the metal atoms are located on the vertices of a tetrahedron 
around each oxygen atom, forming a face-centered-cubic
lattice.~\cite{well84} 
A characteristic feature, as seen  in Fig.~\ref{fig:band}a,
is a linear oxygen-metal-oxygen coordination. 
The calculations for Ag$_{2}$O
are performed with 35 special {\bf k}-points in the irreducible
part of the Brillouin zone (IBZ) of the six atom cell, corresponding
to 1000  points in the full BZ. We used
the same pseudopotentials and energy cut-off (50~Ry) as in our earlier 
works.~\cite{wxli01,wxli02} 

The calculated results, as well 
as available experimental data, are presented in Tab.~\ref{tab:ag2o}. The 
theoretical lattice constant 
of 4.91~\AA\, is 4$\%$ larger than the experimental 
one which is 4.74~\AA\,~\cite{well84} or 4.72\AA\,.~\cite{wyck64} 
The nearest-neighbor distance 
between Ag atoms in Ag$_{2}$O is calculated to be 3.47~\AA\,, which 
is larger than the (theoretical) value in bulk fcc silver 
which is 2.97~\AA\,.
The calculated O-Ag bondlength is slightly shorter than for 
oxygen chemisorbed on the Ag(111)
surface (compare 2.13~\AA\, to 2.15-2.18~\AA\,, depending on the
coverage), and
similar to the O-Ag bondlengths of the $(4\times4)$ structure 
(depicted in Fig.~\ref{fig:commens+4x4} (bottom),  
discussed later and in Ref.~\onlinecite{wxli02})
which are 2.10-2.11 \AA\, 
for the upper O-Ag bonds and
2.14-2.16 \AA\ for the lower O-Ag bonds (involving the surface
Ag atoms).
It is also similar to the O-Ag bond lengths
of the favorable on-surface+sub-surface commensurate oxide
structures (2.10-2.15~\AA\,) reported in Ref.~\onlinecite{wxli02}
[see e.g., the upper O-Ag-O layer of Fig.~\ref{fig:commens+4x4} (top)].

The calculated
heat of formation, obtained as the total-energy difference 
between
di-silver oxide (Ag$_{2}$O) and the sum of the total energies of 
bulk silver (two Ag atoms) and 
the oxygen molecule (one O atom) is 0.092 eV.
The difference to the experimental value of 0.231~eV 
is due to the
significant overbinding of O$_2$, which is
0.566 eV per oxygen atom, 
and the underbinding obtained for bulk Ag and bulk Ag$_{2}$O. 
 This leads to some
cancellation
of errors but it is not
complete.
In table I, the calculated bulk modulus refers to the theoretical
lattice constant. 
To the best of our knowledge, there are no theoretical values
of the heat of formation with which to compare in the literature, nor lattice
constants or bulk moduli.
Both experiment and theory, nevertheless, show that formation of di-silver oxide is an
exothermic process, and the small value is in-line with the low temperature
of decomposition noted above. The heat of formation of di-silver oxide 
is significantly smaller compared to the transition metal
oxides for metals to the left of Ag in the periodic table, e.g. 1.60~eV, 
1.19~eV, and 0.97~eV per oxygen atom (experimental 
data at $T\rightarrow 0$~K and
atmospheric pressure) for 
RuO$_{\rm 2}$, Rh$_{\rm 2}$O$_{\rm 3}$ 
and PdO, respectively.~\cite{crc} 

The total density of states is shown in the right panel of 
Fig.~\ref{fig:band}b. The peak located in the energy range $-$4.0 to 
$-$2.0 eV with respect to Fermi energy arises mainly from the Ag-4$d$ 
states, where there is only a small overlap with the O-2$p$ band, 
which has main (but smaller) peaks in the regions of $-$6
to $-$4~eV and $-$2 to 0~eV.
The states in the former region are largely bonding and those in the
latter are mainly antibonding in character, as determined
by inspecting the spatial distribution of the wave-functions
in these regions. The occupation
of the antibonding states, and small overlap of
O and Ag states, explains the low stability of the oxide. 

The band structure of Ag$_2$O is shown in the left panel of  
Fig.~\ref{fig:band}b. It can be seen that at the $\Gamma$ 
point, the valence and conduction bands overlap, making Ag$_{\rm 2}$O 
metallic, instead of semiconducting with a band gap of 1.3~eV~\cite{tjen90} 
as found in experiments.
The spatial distribution of the wavefunctions in 
the conduction band close to the Fermi energy (investigated for several 
{\bf k}-points), shows that the states are of antibonding nature, involving
a hybridization of Ag-4$d$--O-2$p$- and Ag-4$d$--O-3$s$-like orbitals.
The occupation around $\Gamma$ of these bands could 
contribute to the underestimate of the heat of formation of
Ag$_{2}$O noted above.
We note, however, that this down shift of the conduction band has only a
small effect on the DOS (cf. right panel of Fig.~\ref{fig:band}b at
energy
$\epsilon=0$).
We also performed calculations using the slightly smaller experimental 
lattice constant; this also did not yield a band gap. Using the 
augmented-spherical wave (ASW) method with an extended
basis set and the atomic sphere approximation (ASA) within
the local-density approximation (LDA), Czyzyk {\em et al.} \cite{czyz89} 
calculated the projected density of states
and band structure of Ag$_2$O, where a band gap 0.40~eV was found. 
However, no band gap was obtained by Deb and Chatterjee~\cite{deb98} using
the full-potential linearized augmented plane wave (FP-LAPW) method 
within the local-spin density approximation (LSDA). 
Their reported density of states and band structure are very similar to 
those of our present work. In both of the two above-mentioned 
works, the lattice constant was not stated, but we believe the
experimental one was used because our results calculated at the experimental 
lattice constant agree more closely. 
On increasing the energy cut-off and using
harder pseudopotentials, namely, $r_{s}=r_{d}=2.20$ Bohr 
and $r_{p}=2.10$~Bohr for
silver and $r_{s}=r_{p}=r_{d}=1.2$~Bohr for oxygen, and an energy cut-off 
of 90~Ry,  
no significant change in the results was found
(less than 0.02~eV for the heat of formation, 
0.01~\AA\, for the lattice constant, and 0.001 Mbar for the bulk modulus). 
Thus it appears that Ag$_{2}$O is 
another semiconductor for which the LDA and GGA yield no band gap.
Other materials for which this is the case are, for example, 
Ge,~\cite{staed99} InN,~\cite{stampfl-inn} and ScN.~\cite{stampfl-scn} 

Calculations for Ag$_{2}$O performed using the screened-exchange 
local-density approximation,~\cite{sx-lda}
which affords a more accurate description of the band-gap compared to the LDA
(or GGA),
yield a band-gap of 0.73~eV which
shows that this material is indeed a direct narrow band-gap semiconductor, 
in agreement with experiment.

\section{A\lowercase{g}$_{\rm 2}$O(111) terminations}

\begin{figure}
\scalebox{0.45}{\includegraphics{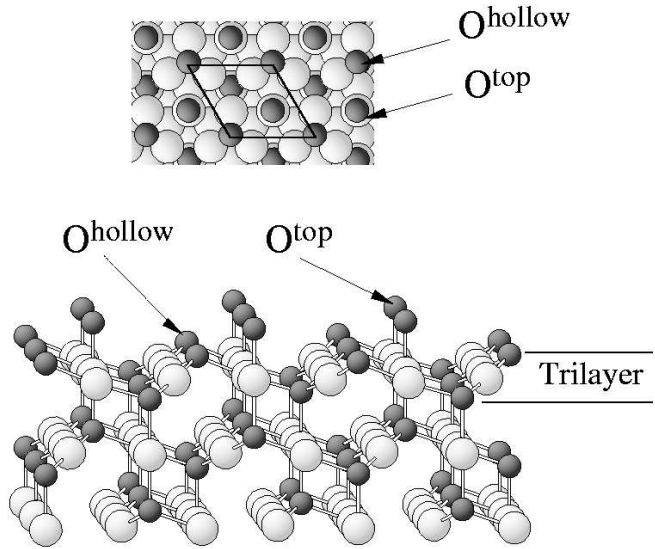}}
\caption{Top (upper) and perspective (lower) views of the atomic geometry
of the Ag$_{\rm 2}$O(111) surface.
The silver atoms are represented by large, pale spheres and the
oxygen atoms by  small, dark spheres.
The hollow and top sites of surface oxygen atoms
are indicated, as is the ``O-Ag-O trilayer'' which contains
two O atoms and four Ag atoms per surface unit cell.}
\label{fig:oxide}
\end{figure}

\begin{figure}
\scalebox{0.45}{\includegraphics{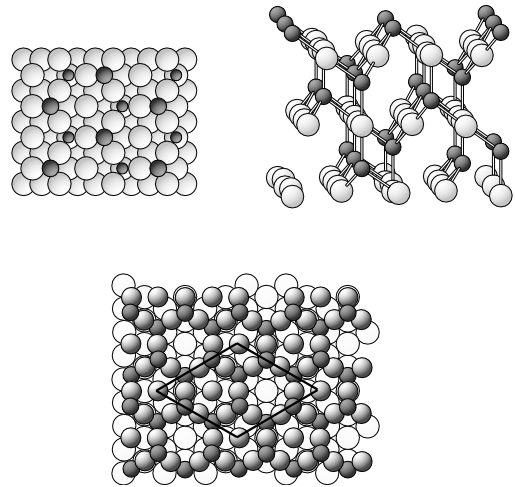}}
\caption{Illustration of the atomic geometry  of an oxide-like
structure commensurate with the Ag(111) surface.
Top view (upper left) and perspective view (upper right) for an O concentration
of 1.25~ML. Ag atoms and O atoms are represented by large pale and
small dark spheres, respectively. The structure is very similar to 
the true oxide (Fig.~\protect\ref{fig:oxide}), but laterally
compressed in comparison.
Lower figure: sketch of the atomic structure proposed in
the STM study of
Ref.~\protect\onlinecite{carl00} for the $(4\times4)$ 
oxygen phase on Ag(111).  The surface unit
cell is indicated. The oxygen atoms are represented by small dark circles, the
uppermost Ag atoms by grey circles, and the intact plane of Ag(111) atoms lying
below the ``O-Ag-O'' trilayer are represented as the 
white circles. Note
that this surface structure is the same as a layer of Ag$_{2}$O(111), except
there is one Ag atom missing in the indicated cell, which would sit
directly on top of the Ag atom in the (111) surface below (which
is energetically unfavorable).
}
\label{fig:commens+4x4}
\end{figure}

In this section, we examine various surface terminations
of Ag$_{2}$O(111). This is interesting for two reasons, firstly in relation
to learning more about oxide surfaces in general through comparison
with existing studies,
and secondly since our earlier works~\cite{wxli01,wxli02} 
for oxygen adsorption at the Ag(111) surface predicted
that for high oxygen concentrations,
thick oxide-like structures form which
contain half a monolayer of O between the 
Ag layers and quarter of a monolayer
on the surface (see e.g., Fig.~\ref{fig:commens+4x4}, upper right).
On inspection of the geometry, we observe that
the structures correspond exactly to that of
the Ag$_2$O(111) surface, but where the positions of the Ag atoms
are
commensurate with those of the underlying Ag(111) substrate, so 
the structure is laterally 
compressed compared to the true oxide [compare Figs. \ref{fig:oxide}
and \ref{fig:commens+4x4} (upper)].
With respect to different terminations, additional
oxygen could adsorb on top of the Ag atoms.
Such sites are called
coordinatively unsaturated sites (``cus''), since compared to in the
bulk, they are lacking their full O-coordination; for the 
present system, the cus Ag atoms are missing one O bond.
Conversely, it
could be favorable that the surface has
a metal-atom termination instead.
We have therefore studied several different terminations, 
and calculated the surface free energy 
as a function of the oxygen chemical potential (cf., 
Eq.~\ref{eq:ochem})
to determine which have the lowest energy.

Figure~\ref{fig:oxide} shows the two surface O atoms:
The hollow site corresponds to the 
bulk stacking and the O atom in the
top site makes this termination oxygen
rich (without the O$^{\rm top}$ atoms the site would be a ``cus'' site).
The silver-terminated surface is that where both of these O atoms are
missing.  The
stoichiometric surface can be thought of as a stacking of O-Ag-O trilayers
as indicated in Fig.~\ref{fig:oxide}, which contains 
two O atoms and four Ag atoms per unit cell of the layer.
We investigated the surfaces 
with oxygen in the hollow site,
and the oxygen-terminated surface with oxygen atoms
in the hollow and top sites. We also tested
the silver terminated surface, as well as 
a termination with oxygen atoms just in the top sites.  

\begin{figure}
\scalebox{0.35}{\includegraphics{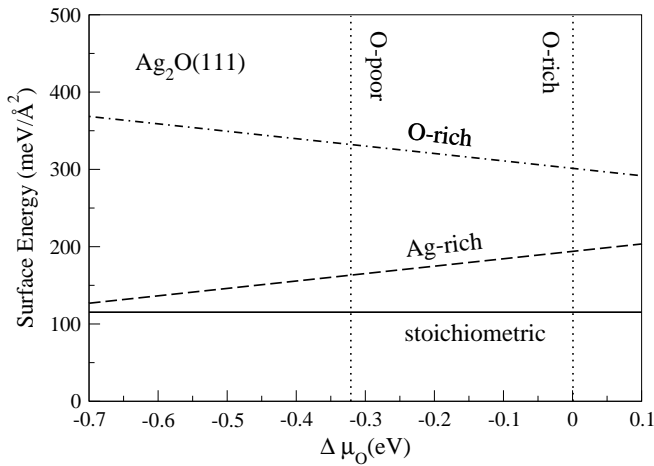}}
\caption{Surface free energies, $\gamma$,
as a function of the O chemical potential,
for the various surface terminations of Ag$_{2}$O(111):
 namely, Ag-terminated (dashed line), oxygen at the hollow
site which corresponds to the stoichiometric
surface (full line), and oxygen at both the hollow and top sites (dot-dashed
line). The left and right vertical lines correspond to the
``oxygen-poor-limit'' and the ``oxygen-rich-limit'', as 
explained in the text, and $\Delta \mu_{\rm O}$ is defined as
$\mu_{\rm O}-1/2E^{\rm total}_{\rm O_2}$, using Eq.~\ref{eq5}
for the latter.  }
\label{fig:oxide-gibbs}
\end{figure}

The surface free energies of the considered surface terminations
are shown in 
Fig.~\ref{fig:oxide-gibbs}. The range of the chemical potential between the
vertical lines corresponds to the (experimental) heat of formation of
Ag$_{2}$O, which is 0.325~eV, i.e. to the range in which bulk di-silver oxide is
stable. It is clear that the surface with oxygen in the hollow site is
the most stable one throughout the {\em whole range} of the chemical
potential. The O terminated surface is very unfavorable. 
This is paralleled by the very large increase in the work
function (by 2.84~eV with respect to the Ag-terminated surface, or
2.50~eV with respect to the stoichiometric surface with O in the
hollow sites).  Such a large work function increase indicates that the
O atoms have a significant negative charge, and so there will
be a considerable O-O repulsion.
The surface terminated with only O atoms in the top site 
is even less favorable than the fully O-covered surface.

The results described above are in contrast to those 
found for other (polar) oxide surfaces, e.g. RuO$_{2}$(110)~\cite{karsten01}
and
$\alpha$-Fe$_{2}$O$_{3}$(0001),~\cite{wang98} where ``full oxygen occupation''
of available sites
becomes favorable for high oxygen chemical potentials. The reason
for this difference
is because here the interaction between oxygen and silver is so weak
that even extremely high values of the
oxygen chemical potential cannot stabilize additional oxygen at the 
polar surface.

In Tab.~\ref{tab:oxide} various physical properties of 
the surfaces are listed.
In particular, it can be seen that the removal energy of
O-hollow is greatest when there is no
additional oxygen at the neighboring top site (compare
3.97~eV with 3.72~eV). Thus, the presence of O in the top
site weakens the bond of the O-hollow atom to the surface, showing     
that there is a repulsive interaction between the O atoms in
the hollow and top sites.
We note that the value of 3.97~eV is significantly stronger than 
3.52~eV as obtained for
O bonded to the clean Ag(111) surface in the fcc site at the 
same on-surface coverage of 0.25~ML as the O-hollow atoms shown
in Fig.~\ref{fig:oxide}.
It is, however, similar to 
the removal energies we found in our earlier work~\cite{wxli02}
for the energetically favorable 
structures involving both on-surface
and sub-surface oxygen and the compressed oxide-like structures, where
the values are in the range 3.90$\sim$4.01~eV. 
These bondstrengths can no longer
be regarded as weak, but are intermediate -- and of similar order as those
identified for the reactive species involved
in certain chemical reactions.~\cite{norskov-uni}

\subsection{Atomic and electronic structure}

The bondlength between oxygen in the hollow site
and the surface silver atom for the energetically favorable 
Ag$_{2}$O(111) structure
is 2.11~\AA\,, which is slightly shorter
than the corresponding bulk value (2.13~\AA\,) noted in Sec.~III. 
It is similar to that of the energetically
favorable compressed oxide-like structures 
(2.10-2.15~\AA\,) [see  Fig.~\ref{fig:commens+4x4} (top)]
which have a very similar local geometry,
but smaller than that for chemisorbed on-surface oxygen
at Ag(111) (2.15-2.18~\AA\,),
as found in our previous works.~\cite{wxli01,wxli02} 
The bondlength between oxygen in the top site
and the silver atom below of 1.96~\AA\,
is significantly shorter because of its low coordination.
Structurally, except for the Ag-terminated surface, where there is 
a large contraction of 10.3~$\%$ of the first metal interlayer spacing 
compared to that of the bulk truncated surface 
(calculated using
the center of the layers),
the relative change in the metal interlayer spacings of the
oxygen-terminated structures is modest, i.e., less than 2.0$~\%$.
For the stoichiometric surface,
the rumpling of the silver atoms in surface metal layer is not negligible;
the magnitude is of the order 0.10~\AA\,.
This is due an outward
movement of the Ag atom in the surface unit cell that is not
bonded to the surface O atoms, i.e.,
the ``cus'' atom.
These relaxations are smaller than those of the ``true 
transition-metal'' oxides (i.e., with incomplete filling of the $d$-band),
e.g. $\alpha$-Fe$_{2}$O$_{3}$(0001)~\cite{wang98} 
and RuO$_{2}$(110),~\cite{karsten01}
which we attribute to
the weaker bonding between silver and
oxygen. Similarly, for adsorption of oxygen on the Ag(111) surface, we found
only very small relaxations of the substrate metal layers~\cite{wxli01}
in contrast to O on Ru(0001)~\cite{stampfl-prb}
where the O-metal bond is notably stronger and oxygen adsorption induces
a significant expansion of the first metal interlayer spacing.
\begin{table}[t!]
\caption{Physical properties of the various surface 
terminations for Ag$_{2}$O(111).
  $E^{\rm removal}_{\rm O-hollow}$ is the removal energy of the oxygen 
  atom which
  occupies the hollow site for the two systems considered
  (cf. Eq.~\ref{eq:ebind}). 
  $\Delta d_{12}$ and $\Delta d_{23}$ are the 
  relative changes in the first two metal
  interlayer spacings with respect to the bulk value (2.83~\AA\,), $b_{\rm
  O-Ag}$ is the bondlength of the surface oxygen atom to the 
  uppermost metal atom, and  $\Phi$ is the work function.
  ``O$^{\rm hollow}$'' is the stoichiometric surface (O$^{\rm top}$ removed,
  cf. Fig.~\ref{fig:oxide}) and ``O$^{\rm full}$'' denotes the
  structure shown in Fig.~\ref{fig:oxide} where both O$^{\rm hollow}$
  and O$^{\rm top}$ atoms are present.} 
  \label{tab:oxide}
\begin{tabular}{l|lllll}
System    & $E^{\rm removal}_{\rm O-hollow}$ & $\Phi$ & $\Delta d_{12}$ & $\Delta d_{23}$ & $b_{\rm O-Ag}$ \\
    & $({\rm eV/atom})$ & $(\rm eV)$& $(\%)$ & $(\%)$ & (\AA\,) 
 \\ \hline 
  Ag-term.       & --   & 4.46 & $-$10.3 & 1.8  & -             \\ 
 O$^{\rm hollow}$  & 3.97  & 5.04 & $-$0.6  & $-$0.7 & 2.11          \\ 
 O$^{\rm full}$  & 3.72  & 7.30 & $-$1.6  & 0.4  & 2.09 (hollow) \\ 
           &       &      &       &      & 1.96 (top)    \\ 
\end{tabular}
\end{table}
\begin{figure}
\scalebox{0.35}{\includegraphics{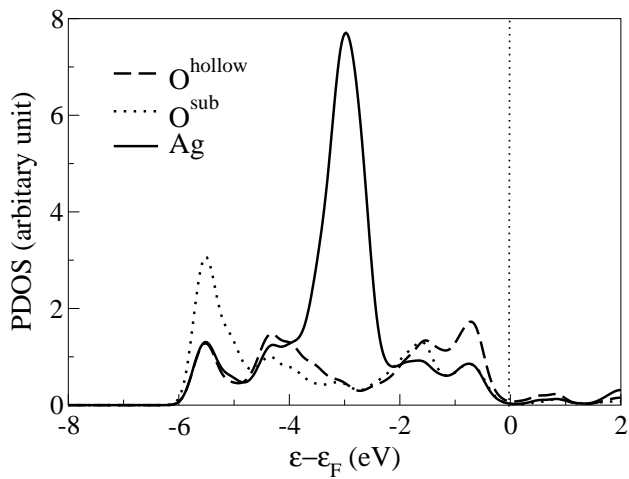}}
\caption{Projected density of states (PDOS)
of the first O-Ag-O trilayer (i.e., the stoichiometric surface): 
The surface oxygen (dashed line) 
which binds to three Ag atoms (it is also called
hollow site oxygen).
  The Ag atom which is bonded to these 
  O$^{\rm hollow}$ atoms (solid line), 
  and the sub-surface oxygen atom  bonded to this Ag atom
  as well as to a 
  single Ag atom in the layer beneath (dot-dashed 
  line) [see Fig.~\ref{fig:oxide}, ignoring
  the O$^{\rm top}$ atoms]. The energy zero is the Fermi energy, as 
  indicated by vertical line. }
\label{fig:oxide-dos}
\end{figure}
The nature of the bonding of the di-silver oxide surface 
is similar to that of the bulk oxide: Oxygen and silver hybridize
mainly via the Ag-4$d$- and O-2$p$-like orbitals 
as shown in Fig.~\ref{fig:oxide-dos}, where
the atom projected density of states of the first O-Ag-O trilayer is shown.
Compared to oxygen chemisorbed on the clean Ag(111) surface
(see Fig.~5 in Ref.~\onlinecite{wxli01}),
the Ag-4$d$ band is significantly narrower,
which is due to
the increased distance between the silver atoms and the weak
hybridization with O-2$p$ states. The PDOS in Fig.~\ref{fig:oxide-dos}
is rather similar to that of the oxide-like structures with same local
bonding as presented in Fig.~12 of 
Ref.~\onlinecite{wxli01}. 
In Fig.~\ref{fig:oxide-dos} two features, one at $-$5.53~eV and 
one at
$-$0.66~eV, with respect to the Fermi energy, can be seen which
are well separated
from the center of the $d$-band. This is in  contrast to the
case of chemisorption on the metal substrate, where there is
only one peak outside of 
the Ag-4$d$ band, for example at $-$1.35~eV for oxygen
in the on-surface fcc-hollow site at coverage
0.25~ML. 
\\

\section{Transition to the Oxide}

Having studied the atomic and electronic structure of 
bulk Ag$_{2}$O and
determined the thermodynamically most favorable surface of Ag$_{2}$O(111), 
we now wish
to investigate how a transition from the oxide-like structures,
with positions of Ag atoms
commensurate with those of Ag(111), and laterally
compressed by 14~\% compared to the true oxide
(as reported in Ref.~\onlinecite{wxli02}), 
to the real oxide structure might occur.
Such a commensurate oxide-like 
structure is depicted in Fig.~\ref{fig:commens+4x4} (top) 
for an O concentration 
of 1.25~ML.
In order to investigate such a transition,
we study the difference of the total energy (per O atom) between
oxide-like films (with commensurate Ag positions to those
of the metal substrate) and corresponding ``real'' (unstrained)
oxide layers as a function of thickness. 
Because the structures can be thought of as the
stacking of O-Ag-O trilayers (see Fig.~\ref{fig:oxide}),
we calculate the energy difference with
increasing numbers of trilayers. 
Then the number of layers at which
the true oxide structure becomes energetically more favorable,
defines a critical thickness for the transition to occur.
In this approach, the coupling between the (strained) oxide-like
film with the underlying metal substrate is not included.
We note that this study may be somewhat ``academic'' since there
is so far no evidence for such compressed oxide structures, 
however,
this structure is a plausible intermediate and perhaps can even
be stabilized under certain
experimental conditions.
Furthermore, additional insight into the O/Ag system can be obtained
by these calculations, and
moreover, when compared to analogous studies performed
for other O/transition-metal systems, in particular
for O/Ru(0001),~\cite{karsten} various trends may be identified.

\begin{figure}
\scalebox{0.45}{\includegraphics{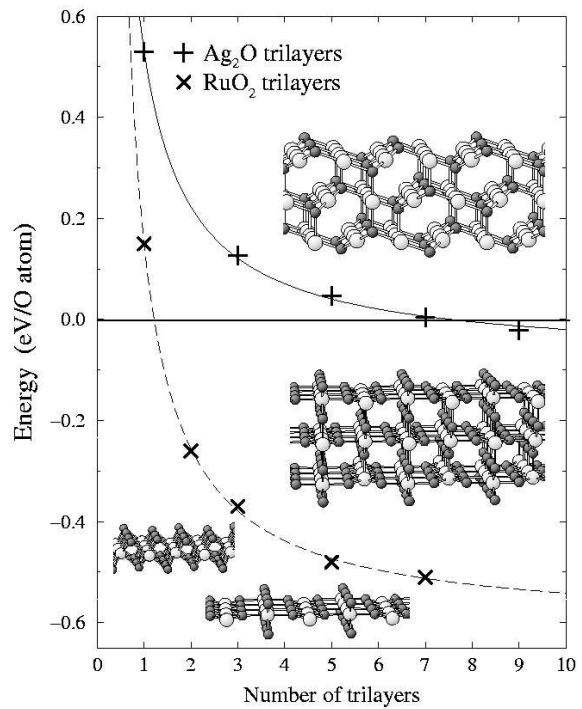}}
\caption{Difference in total energy (per O atom)
 between a stack of O-Ag-O trilayers, 
  commensurate with Ag(111), and that of a corresponding
  oxide layer, as a function of the
  number of trilayers in the stack (denoted by ``+''). 
  The commensurate structures are just like the oxide, only
  laterally compressed by 14~\% and vertically expanded by $\sim$21~\%.
  A negative value indicates that the true oxide structure
  is  energetically more favorable.
  For comparison, analogous results (from Ref.~\protect\onlinecite{karsten})
  are shown for Ru(0001) (denoted by ``{\sf x}'')
  namely, the difference in total energy (per O atom) between a stack
  of O-Ru-O trilayers [commensurate with Ru(0001)]
  and that of corresponding RuO$_{2}$(110) layers as
  a function of the number of layers in the stack.
  The lines connecting the Ag and Ru results (full and dashed lines)
  are guides to the eye.
  Insets illustrate the atomic geometry of 
  the bulk oxide structures where oxygen and metal atoms are represented
  by small dark and large pale circles, respectively.
  For RuO$_{2}$, the O-Ru-O
  trilayer structure is also depicted (lower left), as well as
  its conversion, via an ``accordian-like'' 
  expansion, into a ``true'' RuO$_{2}$(110) layer.  }
\label{fig:transition}
\end{figure}

The results are shown in Fig.~\ref{fig:transition}. 
A negative value indicates that the true oxide structures
are energetically more favorable.
We note that Fig.~\ref{fig:transition} assumes that the number
of O atoms is fixed. However, in many (most) experiments that aim
to study very high coverages, the actual thermodynamic variable may
be $\mu_{\rm O}$.

For one O-Ag-O trilayer, the ``strained'' [commensurate with Ag(111)]
oxide film is significantly 
more favorable than the corresponding true oxide,
as indicated by the relatively large positive value of
0.53~eV, but the preference decreases quickly to 0.13~eV at three
trilayers. On increasing the thickness further, the 
difference in the energy between the two
structures decreases more
slowly, and when the number of trilayers is nine,
the true oxide film is energetically more favorable,
as indicated by the negative value in Fig.~\ref{fig:transition}. 
Because each
trilayer contributes half a monolayer of oxygen, the critical coverage for the
transition from the commensurate oxide-like film to the true oxide film is around
3.5-4.5~ML. As the number of the trilayers approaches infinity, clearly the
curve approaches the value corresponding 
to the energy difference (per O atom) between
the bulk oxide and the ``strained'' [commensurate with Ag(111)] bulk material.

The electronic properties of the
``strained'' [commensurate with Ag(111)] oxide film are very similar to those of
the true oxide
as can be seen from comparison of Fig.~12 in Ref.~\onlinecite{wxli02}
with Fig.~\ref{fig:oxide-dos}. 
This at first may be somewhat surprising since the
lateral dimensions of the
commensurate, oxide-like structures are smaller, by 14~\%. 
We find, however, that much of the associated strain energy is relieved
by expansion of the Ag layers perpendicular to the surface to accommodate
the O atoms in between. Through this relaxation, the lattice volume 
per atom is 92~\% of that of the bulk oxide.

In Fig.~\ref{fig:transition}, analogous results 
for the  phase transition from 
O-Ru-O trilayer structures (commensurate with Ru(0001)) 
to the corresponding real RuO$_{2}$(110)~\cite{karsten}
oxide layers      
are also presented for comparison. 
Bulk ruthenium dioxide has the rutile structure.
The atomic geometry of RuO$_{2}$(110)
is shown as an inset in Fig.~\ref{fig:transition},
as is that of the (commensurate) trilayer and its expansion into a corresponding
true RuO$_{2}$(110) layer (bottom, left).
The commensurate  trilayer structures are, laterally,
15~\% and 35~\% smaller than for RuO$_{2}$(110)
(i.e., in the $\hat{{\bf a}}$ and $\hat{{\bf b}}$ directions
of the rectangular (110) surface unit cell).
In this case
the transformation from commensurate
trilayer to corresponding oxide layer,
involves not only a lateral expansion like for 
the di-silver oxide structures,
but also a ``rotation'' of the O atoms about the Ru atoms, 
as seen from the inset (explained in more detail in 
Ref.~\onlinecite{karsten}).
In particular, every second O-Ru-O component rotates to be horizontal
to the surface, while the alternative 
components rotate so that they are perpendicular to the surface. 
In contrast to the Ag-O system,  the phase
transition into RuO$_2$(110) oxide happens after 
only one O-Ru-O trilayer. Also,   
for Ru(0001), each trilayer contributes two monolayers of
oxygen, so  the critical coverage is around 2-4~ML. 
Because of the considerable
energy gain due to formation of the real RuO$_2$ oxide layers, 
compared to continuous of stacking
of the O-Ru-O trilayers, the formation of the 
oxide structure will
certainly take place once it is kinetically 
possible.

It is interesting to note that the energy differences of
oxygen in the trilayers and the
corresponding bulk metal oxide layers, exhibits
a rather similar dependence on 
the number of layers for both O/Ag and O/Ru, except shifted in energy.
This makes sense since we consider the energy per O atom, 
and as the stack becomes larger, there are more and more 
O atoms in the center of the stack which have atomic environments 
which are similar (and thus similar energies), 
i.e., the affect on the energy due to the surfaces of the layers,
diminishes.
In view of this similarity, together 
with the fact that the heat of formation of e.g., the Rh (1.19~eV)
and Pd (0.97~eV)
oxides lie between the extremes of Ru (1.60~eV) and Ag (0.33~eV) oxides,
we may expect that 
investigations of analogous
transitions from commensurate structures to the bulk
oxide phase for Rh and Pd would exhibit a 
behavior intermediate to those of Ru and Ag.
This would imply the trend 
that the transition from intermediate, commensurate,
two-dimensional oxide films to the true
oxide phase, 
occurs sooner (i.e., for thinner films)
for the elements more to the left in the periodic table,
like Ru, with larger heats of formation, compared
to those towards the right, which have smaller heats of formation like Ag.

From a recent trend study 
on the oxidation at the basal plane of late $4d$ transition metals
(from middle to right in the periodic table, i.e., 
Ru, Rh, Pd, Ag),~\cite{mira01}
it was found that oxygen
starts to enter the sub-surface region only after a certain critical coverage
has been reached on the surface. Namely, at coverages beyond which
the heat of oxygen adsorption
on the surface becomes unfavorable compared
to the heat of formation of the bulk oxide. The 
reported critical coverages
are about 1.0~ML, 1.0~ML, and 0.75~ML, and 0.25~ML for Ru, Rh, Pd, and
Ag, respectively. 
Beyond these critical coverages, onset of bulk oxide formation
was proposed to take place. 
Clearly, however, due to kinetic hindering, bulk oxide formation
may {\em not} be able to occur ``immediately'' and then it may proceed
via the formation of oxide-like layers as described above.

\section{Thermodynamics of the Oxygen-Silver(111) system}

We now turn to investigate the affect
of the temperature and pressure on the stability of the various
oxygen/silver structures,  the energetics and atomic geometries
of which were 
reported in our previous publications;~\cite{wxli01,wxli02}
in particular,
the energies are summarized in Fig.~14 of Ref.~\onlinecite{wxli02}. 
\begin{figure}
\scalebox{0.45}{\includegraphics{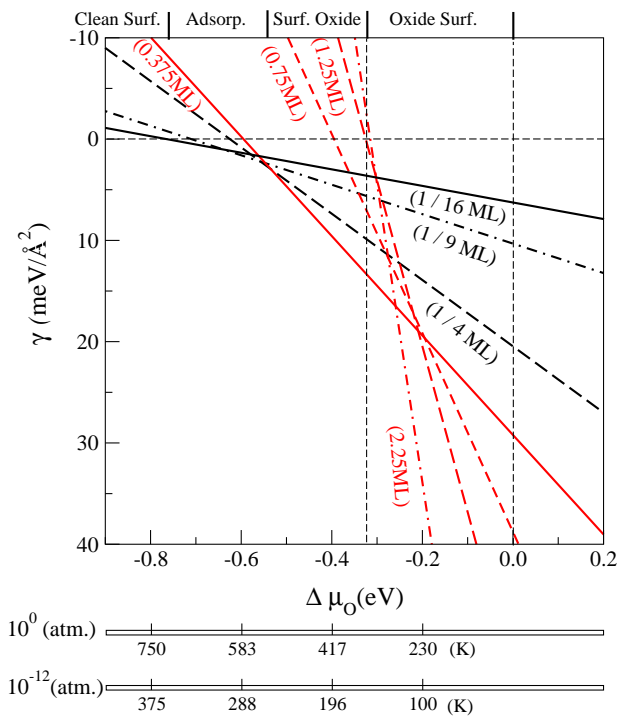}}
\caption{Surface free energies for O at 
  Ag(111) 
  for the various low energy structures as a function of
  the O chemical potential, where $\Delta \mu_{\rm O}$ is defined
  as $\mu_{\rm O}-1/2E^{\rm total}_{\rm O_2}$, using Eq.~\ref{eq5}
  for the latter. 
  On-surface chemisorbed oxygen with
  coverages 1/16~ML, 1/9~ML, and 1/4~ML (labeled by these coverages),
  the $(4\times4)$ structure (accordingly labeled ``0.375~ML''), 
  commensurate oxide-like structures with coverages 0.75~ML,
  1.25~ML, and 2.25~ML (as labeled by the coverages).
  The corresponding temperatures are given for two selected
  pressures (cf. Eq.~\ref{eq:ochem}), one corresponding
  to UHV conditions and the other to the atmospheric conditions of catalysis.
\label{fig:pt}}
\end{figure}

To incorporate the effect of temperature and pressure,
we calculate the Gibbs free energy as a function
of the oxygen chemical potential (cf., Eqs.~\ref{eq-gibbs} and \ref{eq:ochem}) 
similarly to the oxide surfaces, 
except here the reference system is the clean metal substrate.
As discussed above, 
the ``oxygen poor'' limit is that below which
di-silver  oxide is unstable.
For  oxygen adsorption at the Ag(111) surface, however, this limitation
doesn't necessarily exist. 
The results are
shown in Fig.~\ref{fig:pt}, 
where we only show the low energy structures.
The dependence of the energy on $\mu_{\rm O}$, namely the slope
of the line, reflects the O coverage. The higher the coverage, the
steeper the gradient.
We also include the corresponding temperatures for two given pressures, namely,
that which
corresponds to UHV and the other characteristic of the
atmospheric pressure employed in catalysis.

It can be seen that for low values of the oxygen 
chemical potential, i.e. $<-0.78$~eV, the clean Ag(111) surface 
is thermodynamically most stable.
For higher values,
(i.e., in the range of $-$0.78~eV to $-$0.63~eV),
a low coverage (1/16~ML) of oxygen on the surface in fcc-hollow sites
is thermodynamically most stable, and for values 
greater than $-$0.63~eV and less than $-$0.54~eV,
slightly higher coverages of adsorbed O 
(1/9~ML  and 1/4~ML) become most stable.
Increasing the oxygen chemical potential to values greater than $-$0.54~eV
leads to the formation of the reconstructed $(4\times4)$ phase
(labeled ``0.375~ML''),
which is stable up to 
$\Delta \mu_{\rm O}=-0.325$~eV, i.e. to the
``O-poor'' limit (left vertical dashed line), where 
the bulk oxide becomes stable. 
The atomic geometry of 
the $(4\times4)$ phase is illustrated in Fig.~\ref{fig:commens+4x4}
(bottom). 
The atomic structure is very similar to a trilayer of Ag$_{2}$O(111),
except with one Ag in the $(4\times4)$ cell missing.
The basis for this proposed structure (cf. 
Refs.~\onlinecite{carl00,rovi74,camp85,bare95,carl001}) is the 
observation that the area of the 
$(\sqrt{3}\times\sqrt{3})R30^{\circ}$-Ag$_{2}$O(111) surface unit cell
is practically
four times  that of clean $(1\times 1)$-Ag(111); 
the difference 
in the length of the lattice vectors
is only 1.2$\%$  (0.3$\%$ from the experimental lattice
constants). 
Although the mentioned {\em surface unit cells} are 
(practically) commensurate, the
positions of the Ag atoms within the $(\sqrt{3}\times\sqrt{3})R30^{\circ}$
oxide surface cell are,
all except one, incommensurate with those of the Ag(111) surface.

It can be seen that for values of 
$\Delta \mu_{\rm O}$ greater than $-0.54$~eV, 
the commensurate oxide-like structures (see e.g., 
Fig.~\ref{fig:commens+4x4})
(top)
are also quite
favorable (labeled by the O coverages, 0.75, 1.25, 2.25~ML). 
It can be noticed that
the thicker the film, the more stable the structure is 
(i.e., the steeper the curve). 
Clearly, in the limit of an infinitely
thick film, the corresponding
line would be very close to the left vertical one.
This is so because the average O binding energy
in the thick commensurate oxide-like structures
is very similar to that of O in the bulk oxide as
described above.
From Fig.~\ref{fig:pt} 
it can be concluded that the bulk oxide is the
thermodynamically most favorable
structure once the oxygen chemical potential is higher than the ``O-poor''
condition, which is expected. 
However, the kinetics of oxide formation can be very slow and
the thickness of the oxide film will be limited by O and metal
bulk diffusion. Thus, it is unlikely that full thermal
equilibrium of O$_{2}$ with bulk di-silver oxide will be achieved.

\subsection{The O-Ag(111) phase-diagram}
\begin{figure}
\scalebox{0.45}{\includegraphics{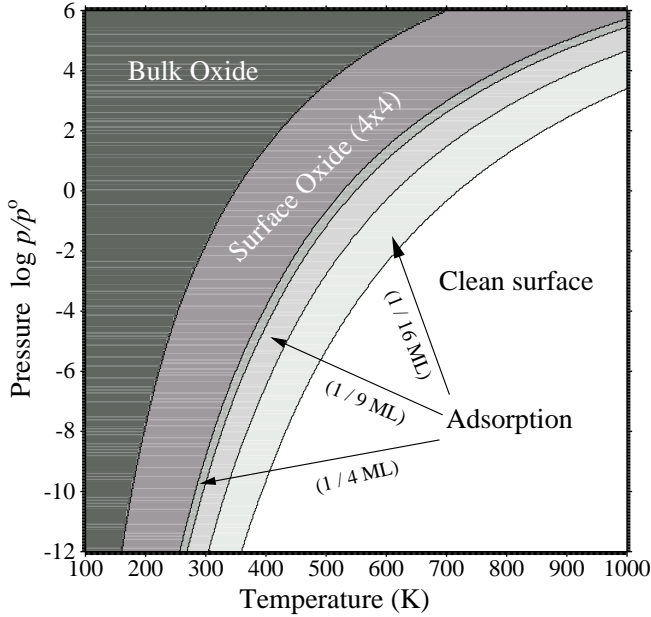}}
\caption{Calculated phase-diagram  for the 
oxygen-Ag(111) system. On the horizontal axis is the
the  temperature in Kelvin, and on the vertical
axis is the pressure. }
\label{fig:pt2}
\end{figure}

The results discussed above can be displayed as a continuous function
of temperature and pressure by constructing a two-dimensional
phase diagram, if we only consider the lowest energy structures.
This has been done in Ref.~\onlinecite{wxli03} and the
results are shown again in Fig.~\ref{fig:pt2}. 
Before giving a discussion, it is
necessary to consider the magnitude of the error bars in the
temperature and pressure. 
An indication can be obtained from knowledge of
the stability of the bulk oxide, i.e.,
that it decomposes at 460~K under atmospheric
pressure. From Figs.~\ref{fig:pt} and \ref{fig:pt2}, the calculations
predict that the bulk oxide decomposes at 
$\sim$350~K at atmospheric pressure, i.e., around 110~K lower than
experiment.
At 460~K, the bulk oxide is 
unstable when the pressure is lower than 10$^{3}$ atm., i.e.,  compared
to 1 atm. as observed experimentally. The reason for
this disparity  
may be due to {\em systematic} errors of the DFT-GGA approach and/or
to
neglect of the entropy contribution. 
In particular, we note that the experimental $-TS$
contribution for bulk silver at 450 K and 1 atm. is $-$0.48 eV per two silver
atoms, and $-$0.70 for Ag$_{2}$O per formula unit.~\cite{crc} Thus, the entropy
contribution from di-silver oxide, compared to silver bulk (the entropy of O$_2$
has already been included in the chemical potential), is greater by 0.22 eV
per two silver atoms, which will shift the left vertical line in
Fig.~\ref{fig:pt} further to the left.
Therefore, in Figs. \ref{fig:pt} and \ref{fig:pt2},
we generally expect that the temperature is underestimated, but the
error less than 110~K, while correspondingly, the
pressure is overestimated, by less than three orders of magnitude. 
For the temperature values quoted below, we give an estimate
of the upper limit in brackets,
that is, 110~K higher than that obtained 
from Fig.~\ref{fig:pt2}.
In spite of these uncertainties, we believe that
the relative stability of the various systems and our
general understanding and conclusions will not be affected.

\begin{figure}
\scalebox{1.10}{\includegraphics{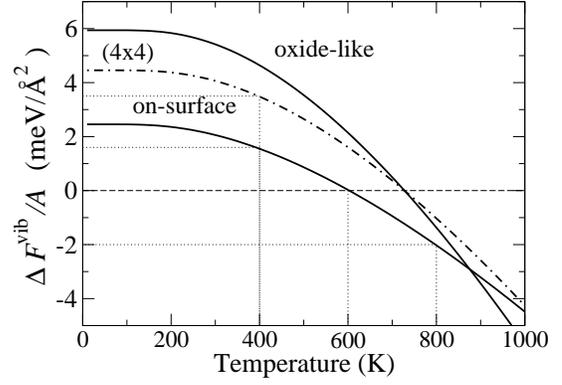}}
\caption{Vibrational contribution 
(oxygen only, see text)
to the Gibbs free energy, $F^{\rm vib}$,
for 
various adsorbate structures on Ag(111).
i) The structure with a coverage of 0.25~ML of oxygen in fcc sites
on the surface, ii) 
for the structure with one O atom in the fcc on-surface site
and one O atom in the sub-surface tetrahedral site 
(labeled ``oxide-like''), and iii) an estimate for the
$(4\times4)$ phase (dot-dashed line).}
\label{fig:vibrations}
\end{figure}

Typically, vibrational
contributions to {\em differences} in the Gibbs free energies
of extended systems exhibit some cancellation.~\cite{karsten01} 
However, when there are additional atomic or molecular species which
are not present in the reference system,
the situation may be different as there will be
no effective cancellation of the contributions of such   
species between surface and reference systems
(as is the case 
using the {\em clean} Ag(111) as reference). 
We investigate this aspect in more detail for the present system; 
in particular, we consider
low coverages of on-surface oxygen and thin oxide-like
structures.
We calculate the difference in
the vibrational and entropic contributions to the Gibbs free
energy, $F^{\rm vib}=E^{\rm vib}-TS^{\rm vib}$, as,
\begin{eqnarray}
\nonumber
\Delta F^{\rm vib}(T,\omega)=\{ \hbar \omega \left( \frac{1}{2} + 
\frac{1}{e^{\beta \hbar \omega} - 1} \right)  
\\
- k_{\rm B}T\left[ \frac{\beta \hbar \omega}{e^{\beta \hbar \omega} - 1} - ln(1
-e^{-\beta \hbar \omega}) \right] \}  \quad ,
\label{eq:vibr}
\end{eqnarray}
where $\beta = 1/k_{B}T$ (see e.g. Ref.~\onlinecite{Scheffler87,karsten01}).
The vibrations which enter Eq.~\ref{eq:vibr} are those
of the adsorbed O atoms.
The calculated perpendicular vibrational frequency for 
0.25~ML of oxygen in fcc sites
on the surface is 50~meV (400~cm$^{-1}$),
and we assume that the lateral vibrations are the same 
(this is a somewhat rough approximation but suffices
for the present discussion, see below).
Thus, in calculating $\Delta F^{\rm vib}(T,\omega)$ we multiply by a factor
of three corresponding to these three modes.
For the thin commensurate oxide-like structure containing one
O atom on the surface in fcc sites and one sub-surface O in
tetrahedral sites (i.e., a total coverage of 0.5~ML), the
calculated perpendicular vibrational frequencies
are 56~meV (455~cm$^{-1}$) and 65~meV (526~cm$^{-1}$),
respectively, for these two species.
Again, assuming each O atom has three modes
equal to the
perpendicular ones, we obtain the results presented in 
Fig.~\ref{fig:vibrations}.

It can be seen 
from Fig.~\ref{fig:vibrations}
that for on-surface O,
for temperatures up to about 800~K, the contribution is a maximum
of about $\pm$2~meV/\AA\,$^{2}$. For lower coverages of on-surface O such
as the 1/16 and 1/9~ML structures, the contribution
will be smaller (assuming the frequency is similar).
For the oxide-like structure at coverage 0.5~ML, the effect
is larger, namely 5-6~meV/\AA\,$^{2}$ for temperatures up to about 400~K, 
thereafter decreasing.
For the $(4\times4)$ thin oxide structure with coverage 0.375~ML,
we expect the vibrational frequencies of the O species will be
similar to those of the structure at 0.5~ML since
the local coordination of the O atoms is similar, as well as the
electronic structure. 
In addition, the vibrational frequency of O in bulk Ag$_{2}$O is
reported to be 66~meV (530~cm$^{-1}$),~\cite{pett}
which is very close to that of the 
O atom in the sub-surface tetrahedral site of the 0.5~ML coverage
structure (65~meV).
Because of the lower O coverage of the $(4\times4)$ structure, the
contribution to the free energy will be smaller than
the 0.5~ML structure by about 0.375/0.5
=0.75, giving a maximum contribution at low temperatures of 
4.5~meV/\AA\,$^{2}$
(dot-dashed line in Fig.~\ref{fig:vibrations}).

For the $(4\times 4)$ structure, in the temperature range 400~K to 730~K,
it gives rise to a destabilization of less than 3.5~meV/\AA\,$^{2}$
(the value at 400~K, which decreases with temperature), which on inspecting
the energy scale of Fig.~\ref{fig:pt} can be seen to be non-negligible.
The relative energy {\em difference}, however between the 
on-surface species and
$(4\times 4)$ structure, which determines the
{\em sequence} of the stable phases, can be seen from Fig.~\ref{fig:vibrations}
to be 
affected by
less than 1.9~meV/\AA\,$^{2}$ ($1.9=3.5 - 1.6$~meV/\AA\,$^{2}$)
for temperatures greater than 400~K, which
will therefore have a small effect.
We note that if the lateral vibrations would be 30~\% greater
than the perpendicular ones for the on-surface
and oxide-like structures considered, this would result
only in a small increase in the contribution to the free
energy, i.e. of  0.5 and 1.2~meV/\AA\,$^{2}$, respectively,
at $T$ close to zero. 

The above analysis may be regarded as qualitative, but it gives an
estimate of the magnitude of the changes in the free energy due to
consideration of vibrational and entropic effects of the extended
systems for the energetically most important O/Ag structures.
Clearly, for other systems with adsorbed species that
have high frequencies such as OH 
(of the order 400~meV),~\cite{sun} the contribution 
to the free energy from vibrational and entropic effects will be
significantly greater and will have to be taken into account
in determination of the phase diagram.

Summarizing the results contained in
Fig.~\ref{fig:pt2}, we obtain the following predictions: 
Under UHV conditions ($p=10^{-12}$~atm.), 
chemisorption of oxygen on Ag(111) with low coverage (less than 0.0625~ML) 
is stable up to about 360~K (incl. corrections: 470~K). This is 
in-line with experimental results~\cite{camp85} where
it has been found that for temperatures
less than 490~K, 0.03~ML of
oxygen can be adsorbed on the Ag(111) surface. 
Under UHV conditions, the calculations also predict the stability
of the $(4\times4)$ phase at the lower temperatures of 160~K 
(incl. corrections: 270~K)
to 260~K (incl. corrections: 370~K). 
This is also consistent with scanning tunneling microscopy (STM) results
obtained under UHV conditions at low (4~K) temperatures~\cite{carl00} where
atomic resolution images of the $(4\times4)$ phase have been achieved. 
The $(4\times4)$ phase, however,
cannot be directly prepared under these conditions as it is known from
experiment that temperatures greater than 400~K
are required in order for the necessary atomic rearrangement take place. 
It is therefore clear that it is the
kinetics that prevent
the reconstruction and formation of this
phase at low pressures and temperatures. 
From Fig.~\ref{fig:pt2}, it can
be seen that the $(4\times4)$ phase should develop into the oxide if the
oxygen pressure is raised or if the temperature
is lower than 160~K (incl. corrections: 270~K). Such transitions, however, may 
be prevented or hindered by kinetics (e.g. diffusion barriers) since
extensive mass transport is required to form the oxide.

At atmospheric pressure ($p=1$~atm.), typical of that used in industry,
the bulk oxide is calculated
to be only stable up to around 350~K (incl. corrections: 460~K). 
From this point of view, it can be concluded that for 
industrial reactions which take place at considerably higher
temperatures, di-silver oxide cannot be the active phase. 
The $(4\times4)$ surface-oxide phase, however, is stable in the
temperature range of 350-530~K (incl. corrections: 460-640~K)
which coincides with that at which 
epoxidation of ethylene occurs. We therefore
propose that such atomic O species are actuating  the reaction.
With regard to
the catalytic reactions that take place at even higher temperatures,
it can be seen that 
only very low coverages of chemisorbed O are stable up to
720~K (incl. corrections: 830~K);
O atoms adsorbed at under-coordinated surface Ag atoms are up to 865~K
(incl. corrections: 975~K) (see below).
Therefore these species could possibly play a role in these reactions.

It is interesting to note that regardless of the pressure, 
Fig.~\ref{fig:pt2} shows that the
{\em ordering} of the stable phases is the same, so that the same phases
should be observable at different pressures if the temperature is adjusted
accordingly.
This indicates 
a way to bridge the pressure gap, i.e., by appropriate choice
of the pressure and temperatures to stay within the same phase. However,
it is clear that sometimes this will not be possible within reasonable
times because of kinetic hindering
for dissociation and sticking of O$_{2}$ (i.e., delivery of sufficient
concentrations of atomic oxygen to the surface), as well as to 
kinetic limitations
for adequate atomic rearrangement.

Experimentally, it has been reported that a ``strongly bound'' oxygen
species (named O$_{\gamma}$)
forms after high temperature ($\sim$900~K) and
atmospheric pressure  exposure of polycrystalline silver, as well
as Ag(111), to an oxygen environment.~\cite{nagy,herein-96}
Temperature programmed desorption spectra show
that O$_{\gamma}$  desorbs at around 900~K and it is proposed by
the authors to
be the active species for the dehydrogenation of methanol~\cite{nagy,bao}
where it diffuses to the surface from bulk sites. 
At the high temperature of $\sim$900~K in an oxygen atmosphere,
silver reconstructs
to form many (111) facets~\cite{nagy,bao} at the surface
and it is speculated that this means that
oxygen diffusion takes place by a so-called
``substitutional diffusion mechanism'' which the authors
argue has higher barriers
than usual interstitial diffusion, and to give rise to the
O$_{2}$ desorption peak at the high temperature of 900~K.
From our calculations, this picture is difficult 
to understand, in particular, how sufficient amounts
of bulk-dissolved oxygen
can be formed.
From our theoretical study, we know that at lower coverage, oxygen will prefer
to stay on the surface, instead of in the sub-surface or bulk 
region, and for high O concentrations, oxide-like structures
are preferred.~\cite{wxli02} 
Furthermore, our results showed that substitutional oxygen adsorption in the
bulk (i.e., O on Ag sites) is energetically unfavorable.~\cite{wxli02}

\subsection{O adsorption at under-coordinated metal sites:Ag vacancies}
The importance of
steps and vacancies as active sites has been well documented, from both 
experimental and theoretical points of view.~\cite{zamb,scha01}
In our previous studies 
we investigated the binding of oxygen atoms on surfaces with pre-existing
vacancies.~\cite{wxli01} 
We found the binding energy  is stronger than on the
(111) terraces, e.g., for 0.25~ML adsorbed in fcc
sites on a $(2\times2)$/Ag(111)  vacancy array
the binding energy is 3.83~eV while it is 3.52~eV on the perfect
terrace, at the same coverage.
A similar behavior can be expected for adsorption at under-coordinated
sites at step edges, as has been shown theoretically for oxygen at
other transition metals, see e.g. Refs.~\onlinecite{norskov-uni} and
\onlinecite{zamb}.
To investigate the affect of pressure and temperature on the stability of such
O species, we calculated the Gibbs free energy of formation.
We find that under atmospheric  pressure, 
the O atoms are stable to around
770~K (incl. corrections: 880~K) on a $(3\times3)$/Ag(111) vacancy array, and even to 865~K (incl. corrections: 975~K)
on a $(2\times2)$/Ag(111) vacancy array.  
The existence and concentration of these species, however, clearly
depends on the number of defect sites
at the surface.
In view of the above, it could be speculated
that the TPD signal of ``O$_{\gamma}$'' is due not to substitutional diffusion
of oxygen atoms from the bulk to the surface as proposed, but due to
O atoms adsorbed
at under-coordinated surface silver sites. 

\subsection{Adsorption of O$_{3}$ at an Ag vacancy}

As mentioned in the introduction,
in the literature, the existence of surface Ag 
vacancies has been proposed to stabilize
a molecular-like ozone (O$_{3}$) species, which has
been suggested to be active 
for the partial oxidation of ethylene.~\cite{avde01}
As reported in Ref.~\onlinecite{wxli02}, 
we calculated the adsorption energy
of such a species on  ($3\times3)$ 
and $(\sqrt{3}\times\sqrt{3})R30^{\circ}$ vacancy arrays, corresponding to
oxygen coverages of 1/3 and 1~ML, respectively. We found that
these species were energetically {\em unfavorable} compared to the
commensurate oxide-like structures described above.
If, however, we assume that Ag vacancies exist, i.e., neglecting
the energy cost to create the vacancy, then the 
{\em binding energy} is the most favorable of all for coverage 1~ML.
We can estimate the concentration of such Ag vacancies using
our calculated values of the vacancy formation energy,
together with an Arrhenius type equation:
For a $(\sqrt{3}\times\sqrt{3})R30^{\circ}$ vacancy array with
a vacancy concentration of 1/3~ML, the formation energy is 0.53~eV;
it is 0.44~eV for a concentration of 0.25~ML, and
0.55~eV for a concentration of 0.11~ML. Considering
a temperature of 550~K, corresponding
to that around which epoxidation of ethylene takes place, this implies
a concentration of thermally induced surface vacancies of less than about
$10^{-5}-10^{-4}$~ML, which is very small.
Furthermore, calculation of the {\em free energy}  shows that 
this species is less stable than the others. 
Our results thus do not support the molecular ozone-like species
as playing an important role,
but point to oxygen atoms of the thin $(4\times4)$ surface-oxide
as being the main species actuating  epoxidation of ethylene.

\subsection{Thick oxide formation at Ag(111)}

As seen from Fig.~\ref{fig:pt}, our calculations predict
that thick oxide-like films can form at Ag(111), but only
for low temperatures [e.g., $\leq$350~K (460~K incl. corrections)
at atmospheric pressure].
The extremely low sticking coefficient of O$_{2}$ at Ag(111),~\cite{gran85}
as well as the found low thermal stability
of these structures, may hinder  their formation.
For example, at high temperatures at which diffusion
of O (or Ag) atoms can effectively occur, the material may become unstable.
Furthermore, 
when using NO$_{2}$ to deliver O atoms to the surface, the
substrate has to be held at temperatures {\em higher} than the temperature
at which di-silver oxide decomposes, in order to desorb the NO molecules.
Using, however, low temperatures and {\em atomic oxygen} or
ozone (which readily dissociates at the surface) instead of
O$_{2}$ (or NO$_{2}$), we predict that thick oxide structures 
could be observed experimentally -- providing that
diffusion barriers can be overcome.

Interestingly,
it was found that di-silver oxide (Ag$_{2}$O) can be prepared using 
a mixture of ozone and O$_2$ or microwave generated atomic oxygen, at room
temperature on {\em polycrystalline} silver foil.~\cite{bhan,jaya00,water,water02}
The oxygen generated in these processes is quite active, for example, small
amounts of 
ozone (5 mol.$\%$ O$_3$ in O$_2$) can 
generate the 
O partial pressure equivalent to that of a pure O$_{2}$ atmosphere
at 10$^{+12}$ atm. at a temperature of 465~K.~\cite{jaya00} 
Actually, in these mixtures of reaction gas, Ag$_2$O
can be stable to temperatures in excess of
773~K.~\cite{water02} In 
the studies of Waterhouse {\em et. al.},~\cite{water} it has 
been found that at room temperature, oxidation of polycrystalline
silver foil is 
improved significantly after the sample has been ``scratched'', 
which generates a high concentration 
of defects which act as the sites for the oxide film
nucleation, highlighting the importance of defects. 
Also diffusion of oxygen and silver atoms is seemingly  improved significantly
due to the non-uniformity of the growth.

We note that subsequent to completion of our studies, very recent
experimental
work reported thick oxide formation at Ag(111)
by using a flux of atomic oxygen,~\cite{longli} thus
supporting our findings.

\section{Conclusion}

Through density-functional theory calculations,
and incorporating the effect of the atmospheric
environment, we obtained the pressure-temperature phase
diagram for the oxygen/Ag(111) system, which reveals important
insights into this system and its function as an oxidation catalyst:
The calculations predict that a thin 
surface-oxide structure is most stable
in the temperature and pressure range of ethylene epoxidation and
we propose it (and possibly other similar structures)
contains the main O species actuating the catalysis.
Low coverages of chemisorbed O become more stable for
higher temperatures and could also play
a role in other oxidation reactions. The only species stable for temperatures
in excess of 720~K (incl. corrections: 830~K), 
are O atoms adsorbed at under-coordinated
surface Ag atoms.
Thick bulk-like di-silver oxide can safely be ruled out as playing
an important role in the technical catalytic
reactions due to the low thermal stability, causing 
decomposition at temperatures lower than those used in the reactors.
They are, however,
the most stable structures for strongly O-rich conditions and low 
temperatures (e.g., up to 330-460~K at atmospheric pressure).
Interestingly, the electronic structure of thick oxide
surfaces are very similar
to that of the thin $(4\times4)$ surface oxide
structure, which raises the question of whether
di-silver oxide might be used as a  {\em low-temperature} oxidation catalyst.

A molecular ozone-like species adsorbed at a surface vacancy,
as had been proposed in the literature as being the active species
for the epoxidation of ethylene, is
energetically unfavorable, as is bulk-dissolved oxygen, which
strongly questions the latters hitherto thought crucial role
in, e.g., the partial oxidation of methanol to formaldehyde.

We also studied the bulk and (111) surface properties 
of Ag$_2$O, the most
stable silver oxide. The interaction of 
oxygen and silver is via a weak hybridization between Ag-4$d$-5$s$ and
O-2$p$ orbitals, with substantial occupation of antibonding
orbitals, which explains its low stability.
Various terminations of the Ag$_2$O(111) surface were investigated
as a function of the oxygen chemical potential
and it is found that the surface with oxygen in the hollow site 
(i.e., the stoichiometric surface)
is most favorable for all of the range of the
O chemical potential considered, while
the Ag-terminated surface and surfaces with additional oxygen
adsorbed on top of exposed ``cus'' surface Ag atoms are unfavorable.
On investigating the energetics of a transition from commensurate
oxide-like structures to the ``real'' (unstrained) oxide, we find that
the transition becomes favorable only at nine
O-Ag-O trilayers, which 
corresponds to $\sim$4.5~ML oxygen.

The approach of ``{\em ab initio} atomistic thermodynamics'',
i.e., of calculating free energies for many
possible atomic geometries and predicting
the lowest energy structures in contact with species in a gas-phase
environment, is a promising approach to investigate
a system under different pressure and temperature conditions,
in particular, from those of typical theoretical surface science, 
right up to those of technical heterogeneous catalysis.
The obtained phase diagram also shows that if one stays within
the region of one phase or along a phase boundary, then
a bridging of the pressure gap should be possible. 
Furthermore, it may be expected that the regions which 
lie close to
the boundaries in the computed surface phase diagram may 
exhibit enhanced thermal fluctuations, i.e., here the dynamics of atomistic
processes may be particularly strong, and this may be particularly relevant
for catalysis under realistic conditions. 
Furthermore,
we note however, that an understanding of a {\em full catalytic cycle},
requires a kinetic modeling that includes all
reactant species and intermediates, which for complex systems such as
those discussed in the present paper,
is not yet possible.
\\


\begin{references}

\bibitem{ertl} 
{\em Handbook on Heterogeneous Catalysis},
Eds. G. Ertl, H. Kn\"{o}zinger and J. Weitkamp,
(Wiley, New York, 1997) Vol.1.

\bibitem{surfsci-500} C. Stampfl, M. V. Ganduglia-Pirovano, K. Reuter, 
M. Scheffler, Surf.  Sci. {\bf 500}, 368 (2002).
     
\bibitem{peden} C. H. F. Peden, in {\em Surface Science of Catalysis:
In situ probes and reaction kinetics}, Eds. D. J. Dwyer and F. M. Hoffmann,
American Chemical Society, Washington DC (1992).


\bibitem{somorjai}
G. A. Somorjai, M. M. Bhasin, J. B. Moffat, and K. I. Tanaka,
Top.  Catal.  {\bf 19}, 143 (2002).

\bibitem{somor2}
K. B. Rider, K. S. Hwang, M. Salmeron, G. A. Somorjai, 
J. Amer. Chem. Soc.  {\bf 124}, 5588 (2002).

\bibitem{freund} H. J. Freund, Surf. Sci.  {\bf 500}, 271 (2002).

\bibitem{ferrer}
K. F. Peters, C. J. Walker, P. Steadman, O. Robach, H. Isern, and S. Ferrer, 
Phys. Rev. Lett. {\bf 86}, 5325 (2001). 

\bibitem{frenken} B. L. M. Hendriksen and J. W. M. Frenken,
Phys. Rev. Lett. {\bf 89}, 046101 (2002).

\bibitem{sant87} R. A. van Santen and H. P. C. E. Kuipers, Adv. Catal.
{\bf 35}, 265 (1987).

\bibitem{rocc01} M. Rocca, L. Vattuone, L. Savio, F. Buatier de Mongeot,
    U. Valbusa, G. Comelli, S. Lizzit, A. Baraldi, G. Paolucci,
    J. A. Groeneveld, and E. J. Baerends, Phys. Rev. B {\bf 63}, 081404(R)
    (2001).    

\bibitem{savi02} L. Savio, L. Vattuone, M. Rocca, F. Buatier de Mongeot,
  G. Comelli, A. Baraldi, S. Lizzit, and G. Paolucci,
  Surf. Sci. {\bf 506}, 213 (2002).

\bibitem{nagy} A. Nagy, G. Mestl, D. Herein, G. Weinberg, E. Kitzelmann,
  R. Schl\"{o}gl, J. Catal. {\bf 182}, 417(1999);
 A. Nagy, G. Mestl, and R. Schl\"{o}gl, J. Catal. {\bf 188},
  58(1999);
 A. Nagy, and G. Mestl, Appl. Catal. A {\bf 188}, 337 (1999). 

\bibitem{bao}  C. Rehren, M. Muhler, X. Bao, R. Schl\"{o}gl, and G.
Ertl,  Z. Phys. Chem. {\bf 174}, 11 (1991);
 X. Bao, M. Muhler, B. Pettinger, R. Schl\"{o}gl, and G.
Ertl,  Catal. Lett. {\bf 22}, 215 (1993);
H. Schubert, U. Tegtmeyer, and  R. Schl\"{o}gl, Catal. Lett. {\bf 28}, 383 
(1994). 



\bibitem{gran85} R. B. Grant and R. M. Lambert, J. Catal. {\bf 92}, 364
  (1985); Surf. Sci. {\bf 146}, 256 (1984).  

\bibitem{van86} R. A. van Santen and C. P. M. de Groot, J. Catal. {\bf 98},
  530 (1986). 

\bibitem{bukh94} V. I. Bukhtiyarov, A. I. Boronin, I. P. Prosvirin, and
  V. I. Savchenko, I. Catal. {\bf 150}, 262 (1994); V. I. Bukhtiyarov, and 
  I. P. Prosvirin and R. I. Kvon, Surf. Sci. {\bf 320}, L47 (1994). 


\bibitem{herein-96}
D. Herein, A. Nagy, H. Schubert, G. Weinberg, E.
Kitzelmann, and R. Schl\"{o}gl, Z. Phys. Chem. {\bf 197}, S67 (1996).



\bibitem{bukh011} V. I. Bukhtiyarov, M. H\"{a}vecker, V. V. Kaichev,
  A. Knop-Gericke, R. W. Mayer and R. Schl\"{o}gl, Catal. Lett.,
  {\bf 74}, 121 (2001).

\bibitem{avde01} V. I. Avdeev and G. M. Zhidomirov, Surf. Sci. {\bf 492}, 137
(2001).


\bibitem{gold} G. C. Bond and D. T. Thompson, Catal. Rev. Sci. Eng. {\bf 41},
319 (1999), and references therein.

\bibitem{gold1} Y. Uchida,  X. Bao, K. Weiss, and
R. Schl\"{o}gl,  Surf. Sci. {\bf 401}, 469 (1998).

\bibitem{gold2} J. Chevrier,  L. Huang, P. Zeppenfeld, and
G. Comsa, Surf. Sci. {\bf 355}, 1 (1996).


\bibitem{copper} Th. Schedel-Niedrig,   Phys. Chem. Chem. Phys.
{\bf 2}, 3473 (2000).

\bibitem{wxli01} W. X. Li, C. Stampfl, and M. Scheffler, 
  Phys. Rev. B {\bf 65}, 075407 (2002).


\bibitem{wxli02} W. X. Li, C. Stampfl, and M. Scheffler, 
  Phys. Rev. B {\bf 67}, 045408 (2003).

\bibitem{wxli03} W. X. Li, C. Stampfl, and M. Scheffler,
submitted to Phys. Rev.Lett.

\bibitem{Weinert86} C.M. Weinert and M. Scheffler,
In: Defects in Semiconductors, edited by H.J. von Bardeleben.
	Mat. Sci. Forum {\bf 10-12}, 25 (1986).

\bibitem{Scheffler87} M. Scheffler,
	In: Physics of Solid Surfaces - 1987,
edited by J. Koukal. Elsevier, Amsterdam 1988, 115 and M. Scheffler
and J. Dabrowski, Phil. Mag. A {\bf 58}, 107 (1988).

\bibitem{Kaxiras87} E. Kaxiras, Y. Bar-Yam, J. D. Joannopoulos,
and K. C. Pandey,
		  Phys. Rev. B {\bf 35}, 9625 (1987).

\bibitem{Qian88} G.-X. Qian, R. Martin, and D.J. Chadi,
	 Phys. Rev. B {\bf 38}, 7649 (1988).

\bibitem{wang98} X. G. Wang, W. Weiss, Sh.K. Shaikhutdinov, M. Ritter,
  M. Petersen, F. Wagner, R. Schl\"ogl, and M. Scheffler,
  Phys. Rev. Lett. {\bf 81}, 1038 (1998). 


  
\bibitem{karsten01} K. Reuter and M. Scheffler, Phys. Rev. B 
  {\bf 65}, 035406 (2002). 
  
\bibitem{fhi98}  M. Bockstedte, A. Kley, J. Neugebauer, and M.
  Scheffler, Comput. Phys. Commun. {\bf 107}, 187 (1997).
  
\bibitem{pbe96} J. P. Perdew, K. Burke, and M. Ernzerhof,
  Phys. Rev. Lett. {\bf 77}, 3865 (1996).
  
\bibitem{white94} J. A. White and D. M. Bird, Phys. Rev. B {\bf 50},
  4954 (1994).     
  
\bibitem{fuch99}  M. Fuchs, M. Scheffler, Comput. Phys. Commun. {\bf
    116}, 1 (1999). 
  
\bibitem{trou91} N. Troullier and J. L. Martins, Phys. Rev. B {\bf 43},
  1993  (1991).  
  
\bibitem{cunningham} S. L. Cunningham, Phys. Rev. B {\bf 10}, 4988 (1974).

\bibitem{jorg92} J. Neugebauer and M. Scheffler, Phys. Rev B {\bf 46},
16067 (1992). 



  
\bibitem{janaf}
  D. R. Stull and H. Prophet, {\em  JANAF Thermochemical Tables}, 2nd ed.,
  U.S. National Bureau of Standards, Washington, D.C. (1971).

\bibitem{crc} {\em CRC Handbook of Chemistry and Physics}, edited by
  R. C. Weast, 55 {\em ed.} Cleveland, Ohio, 1974-1975.

\bibitem{matthias} 
M. Scheffler and C. Stampfl,
in: Handbook of Surface Science, Vol. 2: Electronic Structure, (Eds.)
 K. Horn, M. Scheffler. Elsevier, Amsterdam 1999,
   {\em Theory of adsorption on metal substrates}.

  
\bibitem{wyck64} R. W. G. Wyckoff, {\em Crystal Structures}, Wiley, New York,
  1964. 
  
\bibitem{well84} A. F. Wells, {\em Structural Inorganic Chemistry}, p.1120,
  5th edn., Clarendon, Oxford, 1984.
  
  
  
\bibitem{tjen90} L. H. Tjeng, M. B. J. Meinders, J. van Elp, J. Ghijsen,
  G. A. Sawatzky, and R. L. Johnson, Phys. Rev. B {\bf 41}, 3190 (1990).
  
\bibitem{czyz89} M. T. Czyzyk, R. A. de Groot, G. Dalba, P. Fornasini,
  A. Kisiel, F. Rocca, and E. Burattini, Phys. Rev. B. {\bf 39}, 9831 (1989).
  
\bibitem{deb98} A. Deb and A. K. Chatterjee, J. Phys: Condens. Matter {\bf 10},
  11719 (1998). 

  
\bibitem{staed99}  M. Staedele M. Moukara, J. A. Majewski, P. Vogl, and
  A. G\"{o}rling, Phys. Rev. B {\bf 59}, 10031 (1999).
  
\bibitem{stampfl-inn} C. Stampfl and C. G. Van de Walle, 
Phys. Rev. B {\bf 59}, 5521 (1999); 
C. Stampfl, C. G. Van de Walle, D. Vogel, P. Kruger, and J. Pollmann, 
     Phys. Rev. B {\bf 61}, 7846(R) (2000). 


\bibitem{stampfl-scn}
C. Stampfl, R. Asahi, and A. J. Freeman, Phys. Rev. B {\bf 65}, 161204(R) (2002). 

\bibitem{sx-lda} C. Stampfl (unpublished); R. Asahi, 
W. Mannstadt, and A. J. Freeman,
Phys. Rev. B  {\bf 62}, 2552 (2000).

\bibitem{carl00} C. I. Carlisle, D. A. King, M. L. Bocquet, J.
  Cerd\'{a}, and P. Sautet, Phys. Rev. Lett. {\bf 84}, 3899 (2000).

\bibitem{norskov-uni} J. K. N{\o}rskov, T. Bligaard, A. Logadottir,
S. Bahn, L. B. Hansen, M. Bollinger, and H. Bengaard, and references
therein, J. Catal.  Priority Comm. accepted.



\bibitem{stampfl-prb} 
C. Stampfl and M. Scheffler, Phys. Rev. B {\bf 54}, 2868 (1996).

\bibitem{karsten} K. Reuter, M. V. Ganduglia-Pirovano, C. Stampfl, and
  M. Scheffler, Phys. Rev. B {\bf 65}, 165403 (2002); Chem. Phys. Lett.
   {\bf 352}, 311 (2002). 
  
\bibitem{mira01} M. Todorova, W. X. Li, M. V. Ganduglia-Pirovano, C. Stampfl,
  K. Reuter, and M. Scheffler, Phys. Rev. Lett. {\bf 89}, 096103 (2002). 

\bibitem{rovi74} G. Rovida, F. Pratesi, M. Maglietta, and E. Ferroni,
  Surf. Sci. {\bf 43}, 230 (1974).
  
\bibitem{camp85} C. T. Campbell, Surf. Sci. {\bf 157}, 43 (1985).
  
\bibitem{bare95} S. Bare, K. Griffiths, W. N. Lennard, and H. T. Tang,
  Surf. Sci. {\bf 342}, 185 (1995).
  
  
\bibitem{carl001} C. I. Carlisle, T. Fujimoto, W. S. Sim, and D. A. King,
  Surf. Sci. {\bf 470}, 15 (2000). 
  

\bibitem{pett} B. Pettinger, X. Bao, I. C. Wilcock, M. Muhler, and G. Ertl
    Phys. Rev. Lett. {\bf 72}, 1561 (1994);
	Angewandte Chemie-Intern. Edition in English {\bf 33}, 85 (1994).


\bibitem{sun} Q. Sun, K. Reuter, and M. Scheffler, Phys. Rev. B,
 submitted.

\bibitem{zamb} T. Zambelli, J. Wintterlin, J. Trost and G. Ertl, Science 
  {\bf 273}, 1688 (1996); S. Dahl, A. Logadottir, R. C. Egeberg, J. H. Larsen,
  I. Chorkendorf, E. T\"{o}rnqvist, and J. K. N{\o}rskov,
  Phys. Rev. Lett. {\bf 83}, 1814 (1999); B. Hammer, Phys. Rev. Lett. 
  {\bf 83}, 3681 (1999).
  
\bibitem{scha01} R. Schaub, P. Thostrup, N. Lopez, J. K. N{\o}rskov,
  E. L\ae gsgaard, I. Stensgaard, and F. Besenbacher 
  Phys. Rev. Lett. {\bf 87}, 266104 (2001).
  
\bibitem{bhan} M. K. Bhan, P. K. Nag, G. P. Miller, and J. C. Gregory, 
  J. Vac. Sci. Technol. A {\bf 12}, 699 (1994). 
  
\bibitem{jaya00} K. P. Jayadevan, N. V. Kumar, R. M. Mallya, and K. T. Jacob, 
  J. Mater. Sci. {\bf 35}, 2429 (2000). 
  
\bibitem{water} G. I. N. Waterhouse, G. A. Bowmaker,
  and J. B. Metson, Appl. Surf. Sci. {\bf 183}, 191 (2001).
  
\bibitem{water02} G. I. N. Waterhouse, G. A. Bownmaker and J. B. Metson, 
  Surf. Interface Anal. {\bf 33}, 401 (2002). 
  
\bibitem{longli} Long Li and Judith C. Yang, presented at the MRS
Fall meeting, 2002.
\end{references}
\end{document}